## ■ LA EXPEDICIÓN *PARANÁ RA'ANGA*

[ *ALEJANDRO GANGUI / GRACIELA SILVESTRI / PABLO VENA* ]

**Paraná Ra'anga** *(la figura del Paraná, en guaraní) es el nombre de una Expedición Científico-Cultural que recorrió los ríos de la Plata, Paraná y Paraguay, desde Buenos Aires hasta Asunción, durante el mes de marzo de 2010. El proyecto agrupó a unos cuarenta científicos y artistas de tres países en una lenta y enriquecedora travesía fluvial, poniendo en activo contacto actores de diferentes orígenes y disciplinas que transcurren por separado, en el marco de una experiencia espacio temporal inusual. El proyecto recupera la tradición histórica del viaje como instrumento de conocimiento y colaboración entre las artes y las ciencias, para construir nuevas formas de mirar y comprender el río y sus orillas. Este artículo da cuenta de las motivaciones de este proyecto y de su proyección a futuro.*

La idea inicial de Martín Prieto, director de Centro Cultural Parque de España en Rosario, de remontar el río Paraná siguiendo las huellas de Ulrico Schmidl, primer cronista del Plata, evocaba en nosotros, parte de la futura tripulación, no sólo los relatos maravillosos del ciclo de la conquista, en busca de los metales preciosos, cuando los peces parecían sirenas y las nativas amazonas; también traían a la memoria las novelas de aventuras de la infancia, las de Verne y Salgari, o la refinada tradición de viajeros en lengua inglesa – nuestro William Hudson, Cunninghame Graham y Joseph Conrad. La imagen del viaje en barco por un Paraná que ya no transitaban los viejos vapores de pasajeros, que hacían la carrera hasta Asunción, estaba también teñida de relatos familiares.

Bien sabíamos que no encontraríamos tierras vírgenes, nuevas especies o comunidades no contactadas. No viajaríamos al "corazón de las tinieblas", sino a la Madre de las Ciudades de la cuenca del Plata, Asunción. Transitaríamos por una hidrovía, o por un río cuyo destino cercano era convertirse en hidrovía (poco tiempo después de nuestro regreso, se firmó el convenio para continuar las obras de ingeniería). Pero aunque la región era conocida y explotada, no era percibida ni vivida como la región cultural que alguna vez había sido, cuando el camino del agua era central, y no sólo para las cargas. Uno de los objetivos de la expedición era, pues, hacer pública y visible, una nueva figura del Paraná.

Sin embargo, la expedición tenía un segundo propósito. Se buscaba generar un espacio de encuentros, buscar las afinidades electivas entre integrantes científicos y artistas, amalgamar sus experiencias en los oficios, las artes y las ciencias, para potenciar esta nueva figura (más bien figuras) del Paraná. Días de navegación a paso de hombre, camarotes compartidos, reuniones desalmidonadas (llamadas *convivios*, retomando los convites platónicos, o el "banquete de sabiduría" de Dante), charlas distendidas a la brisa del río, noches recostados en cubierta señalando las estrellas, mostraron a la larga que la interacción y la contaminación de temas y miradas no sólo era bienvenida, sino inevitable.

Un año antes de realizar el proyecto, organizadores y futuros expedicionarios comenzamos a revi-

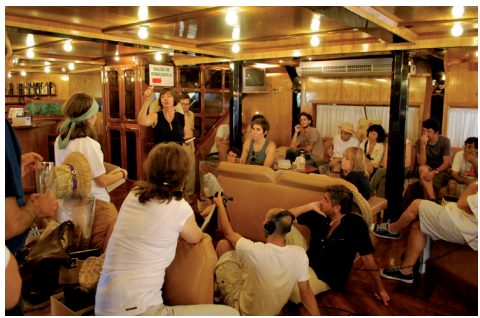

*Reuniones desalmidonadas, bautizadas con el nombre de Convivios, eran el momento de reflexión y discusión de los expedicionarios.*

sar los documentos de los viajes de descubrimiento que habían transitado el Paraná. Las remotas crónicas del mencionado Schmidl, el soldado-cronista que acompañó la expedición de Pedro de Mendoza, participó en la fundación de Asunción del Paraguay y regresó a su patria veinte años más tarde, fueron como dijimos, la inspiración inicial (Box: ***La primera crónica sudamericana***). Los textos del ciclo de la conquista interesan tanto por los hechos comprobados, muchos irónicamente fraseados -"donde ayunó Juan Díaz y los indios comieron"-, como por las leyendas que persisten en la imaginación local y otorgan al territorio una densidad mítica que no se apaga con los años.

Este corpus estaba lejos de ser el único que tuvimos como referencia. En el corazón de esta región del litoral que hoy percibimos fragmentada, los misioneros escribieron el idioma hegemónico, el guaraní, con cuyas palabras se bautizó a la expedición. A los textos de los jesuitas debemos los registros de la historia natural y antropológica en los que se basaron los informes ilustrados. También las utopías de comunidades puras que más tarde fueron intentadas con diversos signos ideológicos.

Pero no fueron los viajes míticos, ni los textos de los jesuitas solitarios, sino los últimos grandes viajes que culminan el ciclo romántico, los que nos alentaron para reunir una tripulación multidisciplinar. El nombre de Alexander Von Humboldt, que no recorrió el sur de Sudamérica, pero presentó una nueva imagen de los dominios hispánicos en una vasta serie de trabajos desplegados en las primeras décadas del siglo XIX, fue la inspiración central desde el punto de vista de la articulación de distintos saberes para describir (para leer, escribir, representar) el rostro de la Tierra en todas sus dimensiones.

La voluntad de sumar diversos aspectos del conocimiento (desde la política hasta el arte, desde la geología hasta la botánica, desde la producción hasta las costumbres), animaba ya los viajes ilustrados, hasta el punto que los portugueses denominaban *viajes filosóficos* a las expediciones científicas que enviaban a sus dominios coloniales, conscientes de que algunas regiones permanecían "tan desconocidas como el primer día del descubrimiento". Como los viajes contemporáneos que envió la corona española, poseían el doble objetivo del conocimiento y el dominio. No era ajeno a esto Alessandro Malaspina, cuya expedición iniciada en 1789, reclutó pintores, antropólogos y naturalistas, bien provistos con instrumentos de precisión náutica, cajas de acuarelas y cámaras oscuras. Uno de los logros más reconocidos de dicha expedición fue el de montar un observatorio astronómico de campaña en Montevideo, que permitió observar el tránsito de Mercurio por delante del disco solar, del día 5 de noviembre de ese año.

Saber y poder se escriben siempre en la misma página: medir y nombrar implican poder sobre lo que se mide y nombra. Este punto fue elocuentemente tematizado por nuestro historietista de a bordo, Pere Joan, mientras el astrofísico del grupo señalaba las estrellas a los menos experimentados: en la cultura occidental las figuras del cielo están determinadas por las historias propias del hemisferio Norte.

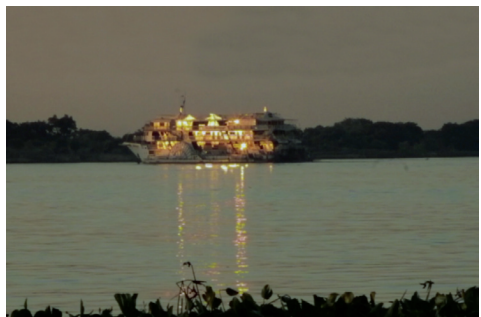
*El crucero Paraguay, visto desde la costa del río.*

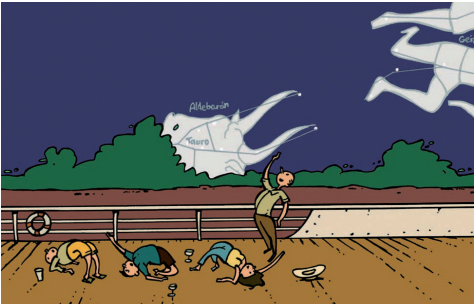

*Medir y nombrar implican poder sobre lo que se mide y nombra, como fue elocuentemente tematizado por nuestro historietista de a bordo, Pere Joan.*

De este matrimonio entre saber y poder emergen muchas discusiones de las últimas décadas, políticas en sentido profundo, de las que intentamos hacernos cargo: ¿es posible *nombrar* de manera plural? ¿Es posible comprender la división de saberes –quiénes tienen derecho a hablar y quiénes no, en las clasificaciones y presupuestos de las ciencias, las técnicas y las artes– *también* como una operación de ordenamiento político del mundo? Pero, ¿es acaso posible suspender todo juicio de verosimilitud (no decimos *verdad*), de eficacia, o de buen o mal funcionamiento?

Como puede imaginarse, tanto antes como durante nuestra expedición contemporánea los debates fueron intensos: la base de acuerdo común –no intentábamos conquistar, sino comprender– solía estrellarse contra las lógicas múltiples de las diversas disciplinas. Es en este punto, de naturaleza epistemológica, en que evocamos una vez más a Humboldt.

En muchos sentidos, somos hijos de la época que se inicia en el siglo XVII, pero alcanza la articulación que todavía manejamos en el siglo XIX, cuando ya el conocimiento científico se estimaba autónomo y de validez universal, íntimamente ligado a los avances tecnológicos con los que se medía el progreso. Pero también en el siglo XIX resultaba un desafío la articulación de los distintos saberes, que iban definiendo sus fronteras; pintar el cuadro de la naturaleza y no sólo desmenuzarla en partes movilizaba la inteligencia de los sabios. Alexander Von Humboldt pretendió armonizar las cartesianas tendencias francesas con el hálito holístico de la *Naturphilosophie*, para superar esta mecánica división de la vida sin desestimar los avances de la ciencia.

Se dirá, pensando en la complejidad del mundo, que estos intentos de reunión estaban –están– condenados a desaparecer. Pero recordemos que Humboldt, *porque* se ocupó de establecer relaciones entre distintas esferas del saber, construyó las bases de la geografía moderna, de la meteorología, incluso de la popular ecología; recreó la narración de la experiencia viajera y estimuló la pintura y la representación del paisaje. No parece así tan ingenuo proponer un camino distinto para pensar el diálogo entre teoría y práctica, experiencia y representación, ciencias y artes.

Más cercano a nuestro tiempo, en viaje por esta misma cuenca fluvial, Claude Levy Strauss no sólo renovó la etnología, también la concepción filosófica del siglo XX: no se mantuvo encerrado en una sola ciencia, y registró con pluma inspirada desde la vida cotidiana de las ciudades hasta la desmesura de la selva, con ojos atentos a la belleza de los pintados rostros caduveos, con sensibilidad alerta para registrar los olores, sabores y sonidos que conformaban paisajes.

**La filosofía del viaje**

Inspirados por estos viajes, pero imposibilitados en el nuevo milenio de encontrar un Humboldt o un Levy Strauss, confiamos en que un colectivo de especialistas -científicos y técnicos, artistas y poetas- podría presentar e inspirar las nuevas figuras del Paraná. Tal esperanza no se fundaba en que, luego de un mes de navegación, se presentaran de inmediato resultados: la inmediatez productiva *no* se corresponde con los viajes. Ellos permanecen como huella en la vida de los viajeros que supieron abrirse a los mundos que visitaban, sin juicio previo ni plan que quedara incólume.

La palabra que usamos: "experiencia" sirvió para afirmar un sentido no productivista. Si esto hubiéramos buscado, bastaba un congreso de especialistas, una serie de videoconferencias, algunas horas de Internet. En alemán, experiencia (*Erfahrung*), remite a un proceso de aprendizaje, a un viaje (*Fahrt*) cuyos frutos se alcanzan al final. También remite a una vivencia que rompe con la trama de convicciones cotidianas. Pero siempre, experiencia remite a los cuerpos: no resulta a distancia. Volvemos a repetir: el viaje de un mes no puede ser repuesto, en sus debates, controversias, acuerdos, comunicaciones aleatorias, y alegrías, en una página web. El proyecto hizo posible reunir todos los sentidos humanos –simbólicos y no simbólicos- bajo el lema que Malaspina adoptó libremente de *La Eneida*: "errante en torno de los objetos miro".

El viaje –todo viaje- es diálogo con el pasado y el futuro. Este diálogo que funda nuestro presente se volvió convocante cuando, en el largo mes que compartimos -más largo que el mes calendario, no sólo por la intensidad de lo inhabitual, sino porque el tiempo se percibe de otra manera en la lentitud del viaje por este río- reconocimos en nuestros debates la larga sombra de debates históricos. El núcleo de nuestras discusiones podría haber sido el de muchos viajeros que se entregaron a la fascinación ambigua de estas tierras.

Meditando acerca de los juicios sobre la "Arcadia perdida" de los jesuitas, el escocés Robert Cunninghame Graham reconoció la contradicción en la que vive nuestra civilización: "la eterna guerra entre los que suponen que el progreso es preferible a una vida tranquila de vana felicidad". Como aquellos viajeros del siglo XIX, nosotros, habitantes de las metrópolis sudamericanas, fuimos sorprendidos por la potencia del Paraná para devorar todo rastro de civilización que no fuera cuidadosamente preservado (cementerios, fábricas y aún ciudades se desvanecieron como las utópicas ciudades de Dios). Para aquellos que nos inclinábamos por el progreso, la experiencia del Paraná medio, sin señal de celular ni Internet, retó con su insólita belleza nuestros sueños transformadores; para quienes adscribían al mito de la felicidad primitiva, no debió resultarles ajena la miseria de los moradores de las orillas castigadas por las inundaciones.

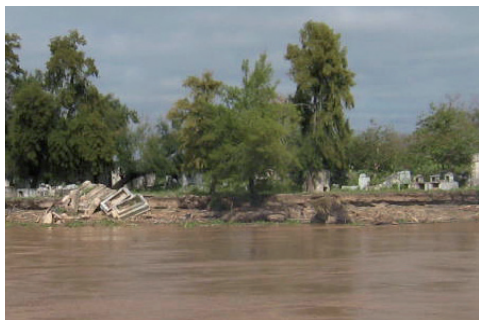

*Cementerios, fábricas y aún ciudades enteras se desvanecen frente a la fuerza del río como las utópicas ciudades de Dios.*

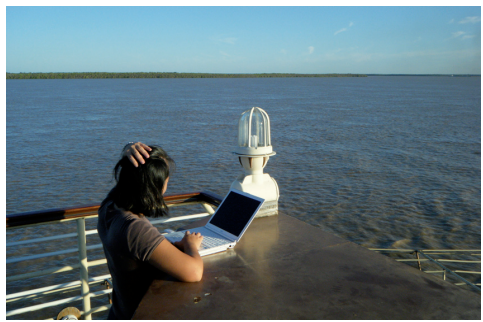

*La experiencia del Paraná medio, sin señal de celular ni Internet. La insólita belleza frente al progreso...*

En todo caso, las conversaciones en cubierta se combinaron con muchos y variados proyectos "en tierra", pues si algún producto imaginamos fue el de establecer relaciones entre quienes habitábamos la misma cuenca, hablábamos la misma lengua y reconocíamos la misma historia. Esta combinación no podría ser efectiva desde el escritorio. Se trataba de promover algo que permanece ausente en las consideraciones de las ciencias exactas y humanas, y que está en vías de desaparición en las artes, la música, las letras: el cara a cara entre las personas que escriben, miden, registran; entre personas que actúan, viven y esperan.

**Hacia figuras en transformación**

Como lo podíamos anticipar, la actividad a bordo fue muy variada. El segundo piso del Crucero Paraguay, cerca de la barra, era el lugar preferido de los artistas. Entre ellos, mezclados, tres arquitectos, una maestranda en demografía, dos ingenieros y un geógrafo, la "verdadera tripulación" del barco esgrimiendo una cámara común, una historiadora del arte con manos sucias de crayón, una bióloga siguiendo el paso de un camalote, y un poco más alejado un escritor con una libretita de notas en sus manos. A la sombra del Sol, varios expedicionarios desplegaban su artillería: acuarelas, pasteles, témperas, chablon, con sus solventes y trapos. Otros leían y dialogaban sobre el paisaje, sobre las poblaciones y la urbanización de las orillas del río.

Aquellos filmaban, estos escribían, un pequeño grupo tomaba una guitarra y tarareaba una canción, quizás una conocida chamarrita litoraleña, quizás las primeras notas de una nueva composición inspirada en la costa entrerriana de estribor. Algo más lejos se oía un piano eléctrico donde se ensayaban acordes operísticos. Se discutía sobre las costumbres musicales de los pueblos que atravesaba la hidrovía, sobre la economía de la región, sobre el arte comprometido con la preservación del río. También la cuestión culinaria era tema –más que de discusión, de degustación.

Entre los expedicionarios se hallaba Ignacio Fontclara, quien se autodefinía como panadero, luego pastelero, y más tarde cocinero, hortelano, y todo antecedido de "aprendiz de…", pero que para sus compañeros se convirtió en una suerte de filósofo de la alimentación, secundado por Emilio Nasser, becario cocinero y fotógrafo. En su mochila Ignacio traía, bien conservada, su herramienta "viva" de trabajo: la "masa madre", un cultivo de las levaduras naturales de los cereales y de las bacterias presentes en el aire. Como una antigua tradición, los panaderos de oficio conservan, alimentan y dejan fermentar a esa masa madre con cuidado, agregándole harina y agua. El traslado de la masa hace que incorpore las características del medio ambiente donde se halla, dándole a los alimentos formados con ella un toque diferente -e irrepetible- de acuerdo a su historia previa. Las explicaciones de Ignacio se complementaban con las del etnomusicólogo Guillermo Sequera, y las del lingüista Bartomeu Meliá. En la labor de cocina estaba magistralmente resumido el objetivo de la expedición: el cruce de diversas disciplinas –la geografía, la antropología, la historia de las costumbres, la bo-

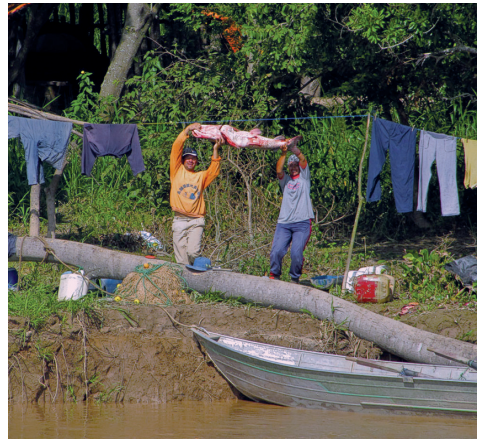

...aunque también la precariedad de los moradores de las orillas castigadas por las inundaciones.

tánica, la ecología–, resultando en hallazgos que podían convocar todos los sentidos.

Resultaría ocioso describir la multitud de proyectos, individuales y colectivos, que surgieron en el barco. Algunos, a algo más de dos años de ese viaje insólito, han derivado en productos específicos, como la musicalización por Jorge Fandermole de unos versos en Guaraní de *Cielito marangatu* (Cielito el bondadoso) con adaptación de Dionisio Arzamendia y Sequera (Box: *El Cielito Guaraní*), o el libro de historietas del artista mallorquí Pere Joan. Otros se abrieron a nuevos intercambios luego del fin de la expedición –desde los abiertos entonces, como los que la diseñadora y arquitecta Claudia Tchira y el pintor Fernando Bedoya establecieron con el centro cultural del barrio Toba, en Rosario–, como los que se impulsaron más adelante, articulándose con otros proyectos sobre el eje del Paraná, por ejemplo, los trabajos de Gabriela Siracusano con las imágenes de La Paz (Entre Ríos) o los estudios sobre infraestructura y paisaje en Sudamérica, que ingenieros y arquitectos de a bordo continuaron desarrollando.

Durante el 2011, los organizadores decidieron reunir este universo múltiple en una muestra itinerante y en un libro. La muestra *Itinerancia 2011-2012* iniciaría su viaje en sentido inverso: desde Asunción hasta Buenos Aires –para recalar, finalmente, en Madrid. Se inauguró el 21 de octubre de 2011, en el marco de los festejos del Bicentenario paraguayo, curada por María Teresa Constantin, y actualmente se encuentra en viaje hacia los centros urbanos que la expedición fue tocando. Esta exhibición, por supuesto, se encuentra en permanente ampliación y viene acompañada de actividades variadas, como debates, conciertos y talleres. En cuanto al libro, que contó con el diseño de Juan Lo Bianco, fue titulado *Paraná Ra'anga, un viaje filosófico*, haciéndose eco de la vieja tradición ibérica de los viajes ilustrados. Como en las viejas crónicas, el núcleo de este documento está compuesto por un diario de bitácora, montaje de los diarios de Martín Prieto, Daniel García Helder y María Moreno, los escritores del barco, ampliamente ilustrado con el material de los fotógrafos y artistas visuales. Allí se incluyen también presentaciones geográficas e históricas de la región, destinadas a un público amplio, no necesariamente sudamericano, y artículos y ensayos acerca de los proyectos en curso. No es secundario que se haya decidido invitar en esta empresa a dos reconocidos curadores y diseñadores que no habían participado de la expedición: la idea era la de trasmitir la experiencia, no atesorarla sólo para la pequeña cofradía viajera.

Y así las figuras del Paraná continúan su transformación, en el entusiasmo de otros que reescriben lo trasmitido. Porque finalmente, lo que se aprende en un viaje en barco –en un viaje inusual como este, que suspende el trabajo cotidiano, las normas habituales, para entrar en un mundo acuático cuya lenta temporalidad transforma la propia percepción del espacio– es que, como dice el lema de Pompeyo, tantas veces repetido, *navigare necesse est.*

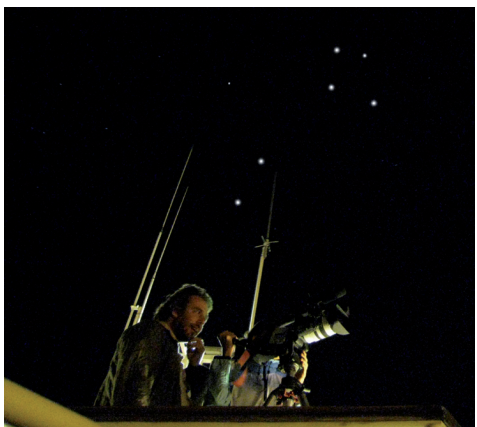

*Las figuras del Paraná también se plasman en el cielo y la noche fue el escenario predilecto de muchos expedicionarios. (A la derecha se llega a ver la Cruz del Sur y, entre los dos mástiles, aparecen alfa y beta del Centauro.)*

## La primera crónica sudamericana

En 1534 el soldado alemán Ulrico Schmidl partió del puerto de Cádiz rumbo a América, como parte de la tripulación de la expedición de Don Pedro de Mendoza al Río de la Plata, quedándose en territorio americano por casi veinte años, ocupando en la colonia cargos de jerarquía menor, hasta 1552, cuando regresa a Alemania. En ese entonces en Europa Gian Batista Ramusio había publicado tres volúmenes de su *Delle Navegazione e Viaggi*, cuya enorme repercusión marcó la conciencia europea acerca del giro que producían en la historia del continente y de la Humanidad los grandes viajes y descubrimientos: la conciencia acerca de la forma que iba tomando el mundo después de Colón y de Gutenberg.

Schmidl, que no era escritor ni amanuense, motivado por la expectativa que creaban los libros de viajes, decidió escribir el suyo propio, que se publicó en Frankfurt por primera vez en 1567, bajo el título Viaje al Río de la Plata. El libro fue escrito en una especie de lengua franca en la que se mezclaban el alemán –la lengua base de su escritura- con hispanismos e indigenismos. Y si bien esa decisión debe haber provocado desconcierto entre sus lectores contemporáneos -para quienes el libro pasó casi desapercibido- en perspectiva le da al libro el valor agregado de un documento que respaldó investigaciones de lingüistas y antropólogos. Sin embargo, durante muchísimos años la obra de Schmidl fue despreciada como fuente histórica y como literatura. En cuanto a lo primero, debido tanto a sus descripciones idénticas de territorios o tribus diferentes, como a sus desvíos fantásticos, propios de la literatura de viajes –como el relato sobre su llegada a la tribu de las amazonas, las mujeres de un solo pecho que vivían en una isla. Y en cuanto a la literatura, porque la llaneza de su estilo muchas veces no logra ser disimulada por la realmente singular aventura que está narrando.

El paso del tiempo devolvió a la obra del alemán esas dos condiciones: históricamente, la relación sigue teniendo el valor de una fuente documental de primera mano sobre algunos de los episodios sucedidos durante la expedición de Mendoza al Río de la Plata –la fundación de Buenos Aires, la llegada al fuerte de Asunción, las disputas entre adelantados- narradas siempre con curiosidad y vigor. Y esa misma curiosidad es la que hoy nos permite tener una primera imagen, a caballo entre la historia, la etnografía y la literatura, de las poblaciones de charrúas, querandíes, guaraníes, chaná-timbúes, carios, mapenis, mocoretás, agaces, surucuces, jerús, corondás, entre otros: de su aspecto físico, de sus costumbres, de su alimentación, de sus maneras de vestirse, cazar, pescar, trabajar la tierra, guerrear.

Literariamente, pese a su precariedad compositiva, hay dos condiciones que otorgan lozanía al texto de Schmidl. Una es la del uso de la comparación como recurso retórico privilegiado. Pero no una comparación poética, literaria, sino una comparación pedagógica, como un instrumento de comprensión para los europeos del nuevo mundo, de modo que los querandíes, en tanto nómades, son comparados con los gitanos, las boleadoras son descritas "como las plomadas que usamos en Alemania", la raíz de batata se parece a la manzana, la mandioca a la castaña, etc. Y esa metáfora pedagógica que tiende a reunir campos semánticos alejados entre sí, tiene un enorme valor como antecedente de las relaciones establecidas entre europeos y americanos y más precisamente, entre españoles e hispanoamericanos a partir de los flujos y reflujos migratorios sucedidos desde fines del siglo XIX en adelante, de manera ininterrumpida: la búsqueda de lo común en lo diferente.

La otra condición, de la que su obra es menos consciente pero que la historia de la literatura ha sabido valorar, es que *Viaje al Río de la Plata* es la primera construcción simbólica de un escenario fluvial, que va de Buenos Aires a Asunción. Ese mismo escenario, retomado después en las obras de ficción de algunos de los mayores escritores argentinos y paraguayos –Juan José Saer, Horacio Quiroga, Augusto Roa Bastos, entre muchos otros- además de pintores, cineastas y cronistas, le otorgan, retrospectivamente, a la obra de Schimdl, un lugar privilegiado en la historia del arte local, lugar del que pueden jactarse muy pocas obras del artísticamente rudo período colonial rioplatense.

*Martín Prieto, escritor, Director del Centro Cultural Parque de España (Rosario).*

*[ Alejandro Gangui / Graciela Silvestri / Pablo Vena ]*

**El Cielito Guaraní**

El Cielito es un estilo musical de gran difusión en la época de las guerras de la Independencia en el Río de la Plata. Bartolomé Hidalgo y otros poetas insurgentes adoptan la forma cielito (en poesía, música y danza) como símbolos patrios y soberanos. En Paraguay, durante la Guerra de la Triple Alianza, Francisco Solano López encomienda a Natalicio Talavera (joven poeta paraguayo) dirigir los diarios de campaña, impresos en la retaguardia guerrera. El poeta, que muere de tifus y en batalla en Paso Puku, a los 27 años de edad, escribió este tema. Dionisio Arzamendia y Guillermo Sequera lo actualizaron. Este último, acompañado por la guitarra de Jorge Fandermole, lo cantó en el barco, constituyéndose como el "himno" de la expedición.

**Cielito marangatu**

*Letra: Natalicio Talavera, poeta guaraní*
*Adaptación y Música: Dionisio Arzamendia (músico, arpista)*
*Guillermo Sequera (etnomusicólogo)*

| | |
|---|---|
| Allá viene Cielito | *Amóuina ou Cielito* |
| trayendo sobre sus hombros | *Ijapyri ogueru* |
| A un niño descalzo | *Mita'i pynandymi* |
| Su madre lo siente venir | *Isymíme oñandu* |
| | |
| La madre sufriendo | *Isymi hasy katuva* |
| Hace mucho de tuberculosis | *Aretema ihu'u* |
| Medicamentos no tiene | *Pohâmi ndoguerekóigui* |
| largo dolor | *Hasy po'i ndajeku* |
| | |
| Al llegar hasta el ranchito | *Oguâhêvo oga guype* |
| Desde dentro se escucha: | *Kotypýgui ñahendu* |
| Bienvenido Cielito | *Eikemíkena Cielito* |
| Traes a mi hijito? | *Rereúiko che memby* |
| | |
| Hace tiempo Cielito | *Aretéma ko Cielito* |
| Se fué Lakú | *Che reja hâguê Laku* |
| en busca de alimentos | *Oho vaekue ohekávo* |
| para mi | *Chéve guâra tembi'u* |
| | |
| Ya se fue Cielito | *Ohokévoma Cielito* |
| Y el niño contento: | *Mita'ípe oipopyhy* |
| Cielito le dice | *Ha he'i chupe ra'yto* |
| Cura y cuida a tu mamita | *Tokuerákena nde sy* |

**Bibliografía**

Schmidl U., *Viaje al Río de la Plata*, 1534-1554, notas bibliográficas y biográficas por Bartolomé Mitre, prólogo, traducciones y anotaciones por Samuel Alejandro Lafone Quevedo, Alicante: Biblioteca Virtual Miguel de Cervantes, 2001, http://www.cervantesvirtual.com/obra/viaje-al-rio-de-la-plata-1534-1554/ .

Prieto, M. y Silvestri G. (editores), *Paraná Ra'anga, un viaje filosófico*, Rosario: Centro Cultural Parque de España, 2011.

Más imágenes de la Expedición y de los Cielos del Paraná Ra'anga, en https://picasaweb.google.com/algangui/ .

**Agradecimientos**